\documentclass[12pt,a4paper]{article}

\usepackage{latexsym}
\usepackage{amsmath}
\usepackage{amsfonts}
\usepackage{amssymb}

\newcommand{\be}{\begin{equation}}
\newcommand{\ee}{\end{equation}}
\newcommand{\bea}{\begin{eqnarray}}
\newcommand{\eea}{\end{eqnarray}}

\def\ad{{\mathrm{ad}}}                  %
\def\G{{\cal G}}                        %
\def\M{{\cal M}}                        %
\def\H{{\cal H}}                        %
\def\A{{\cal A}}                        %
\def\cO{{\cal O}}                       %
\def\ri{{\mathrm{i}}}                   %
\def\rr{{\mathrm{r}}}                   %
\def\cR{{\cal R}}                       %
\def\cL{{\cal L}}                       %
\def\bR{{\mathbb R}}                    %
\def\bC{{\mathbb C}}                    %
\def\1{{\mbox{\boldmath $1$}}}          %
\def\tr{\mathrm{tr\,}}                  %
\def\diag{\mathrm{diag}}                %
\def\0{{\mbox{\boldmath $0$}}}          %
\def\y{{\cal Y}}                        %
\def\ext{\mathrm{ext}}                  %
\def\red{\mathrm{red}}                 %
\begin{document}

\vspace*{0.5cm}
\begin{center}
{\Large \bf A class of Calogero type reductions of free motion
on a simple Lie group}
\end{center}

\vspace{0.2cm}

\begin{center}
L. FEH\'ER${}^{a}$ and B.G. PUSZTAI${}^b$ \\

\bigskip

${}^a$Department of Theoretical Physics, MTA  KFKI RMKI\\
1525 Budapest 114, P.O.B. 49,  Hungary, and\\
Department of Theoretical Physics, University of Szeged\\
Tisza Lajos krt 84-86, H-6720 Szeged, Hungary\\
e-mail: lfeher@rmki.kfki.hu

\bigskip
${}^b$Centre de recherches math\'ematiques, Universit\'e de Montr\'eal\\
C.P. 6128, succ. centre ville, Montr\'eal, Qu\'ebec, Canada H3C 3J7, and\\
Department of Mathematics and Statistics, Concordia University\\
1455 de Maisonneuve Blvd. West, Montr\'eal, Qu\'ebec, Canada H3G 1M8\\
e-mail: pusztai@CRM.UMontreal.CA

\end{center}

\vspace{0.2cm}

\begin{abstract}
The reductions of the free geodesic motion
on a non-compact simple Lie group $G$ based on the
 $G_+ \times G_+$ symmetry given by left- and right-multiplications for
a maximal compact subgroup $G_+ \subset G$ are investigated.
At generic values of the momentum map this leads to (new) spin Calogero type
models.  At some special values the `spin' degrees of freedom are
absent and we obtain the standard $BC_n$ Sutherland model {\em with
three independent coupling constants} from $SU(n+1,n)$ and from $SU(n,n)$.
This generalization of the Olshanetsky-Perelomov derivation
of
the $BC_n$ model with two independent coupling constants from
the geodesics on $G/G_+$ with $G=SU(n+1,n)$ relies on
fixing the right-handed momentum to a non-zero character of $G_+$.
The reductions considered permit further generalizations and work at
the quantized level, too, for
non-compact as well as for compact $G$.

\bigskip
\noindent
\textbf{Mathematics Subject Classifications (2000).} 37J35, 53D20, 17B80, 70G65.

\bigskip
\noindent
\textbf{Key words.} integrable systems, Hamiltonian reduction, spin Calogero models,
$BC_n$ Sutherland model.

\end{abstract}

\newpage

\section{Introduction}
\setcounter{equation}{0}

The `Calogero type' integrable models of interacting particles on the line
are  interesting on account of their physical applications and
relationships to important fields of  mathematics.
Generalizations of the original model \cite{Cal}  can be associated with root systems
in correspondence
with various admissible interaction potentials and possible couplings
to internal `spin' degrees of freedom and to external fields.
The richness of these models is demonstrated by the
growing number of reviews devoted to
them \cite{OPRep1,OPRep2,Per,Heck,Nekr,dHP,Diej,SutWSci,Finkel,Eting,Poly,AniJev}.
One of the basic models of the family is the hyperbolic
$BC_n$ Sutherland model defined classically
by the Hamiltonian
\be
\H_{BC_n} = \frac{1}{2}\sum_{k=1}^n p_k^2
+\sum_{1\leq j<k\leq n}\bigl( \frac{g^2}{\sinh^2(q^j-q^k)}
+ \frac{g^2}{\sinh^2(q^j+q^k)}\bigr)
+ \sum_{k=1}^n\bigl(\frac{g_1^2}{\sinh^2(q^k)}
+ \frac{g_2^2}{\sinh^2(2q^k)}\bigr)
\label{1.1}\ee
with arbitrary  coupling constants $g, g_1, g_2$.
Olshanetsky and Perelomov \cite{OPInv,OPCim,OPRep1} showed
that this model can be viewed as a `projection' of
the geodesic  system on the symmetric space $SU(n+1,n)/(S (U(n+1) \times U(n))$  if
the coupling constants obey the quadratic relation
$g_1^2 -2g^2 +\sqrt{2} gg_2=0$.
For arbitrary coupling constants, classical and quantum
solvability of the model was  established by means of
different, rather algebraic, methods \cite{InoMes,Opd,OshSek,Sas}.

Since Hamiltonian reduction is a very effective and
general approach to integrable systems, it
would be interesting  to lift the above quadratic relation of Olshanetsky and Perelomov
sticking to this method. Motivated partly by this problem, recently we undertook a systematic
study of reductions of the free geodesic motion on Riemannian symmetric spaces, which
led to new spin Calogero models as well as to an understanding of the
geometric origin of the quadratic relation \cite{FP}.
Here, we extend this work by going one stage up and explore
the reductions
of the geodesic system defined on the isometry group of the symmetric space.
We shall demonstrate that
 the classical $BC_n$ model (\ref{1.1}) with three independent coupling constants can be obtained
by Hamiltonian reduction in this extended framework.

The geodesic system on a symmetric space, realized as a coset space $G/G_+$,
is a reduction of the
geodesic system on the isometry group $G$, belonging to the zero value of the momentum map for
the action
of the little group $G_+$ on $T^*G$ generated by right-multiplications.
This system then can be reduced to spin
Calogero models using the residual symmetry generated by the
left-multiplications associated with $G_+$.
It is clear that
more general reduced systems result if one fixes the right-handed momentum to
some non-zero value.
First, we shall describe the most general reductions of $T^*G$ that rely
on the action of $G_+ \times G_+$ through left- and right multiplications.
In fact, one obtains (new) spin Calogero type models in general, with the
spin degrees of freedom restricted to a trivial one-point space in certain very special cases.
Second, we  observe that if the space of spin degrees of freedom
is trivial for the zero
value of the `right-handed' momentum map,  then this feature can be ensured also for any
non-zero character (one-point coadjoint orbit) of $G_+$.
Taking advantage of this observation,
we can derive the $BC_n$ model with three
independent coupling constants from the geodesic motion on $SU(m,n)$ both for $m=n$ and for $m=n+1$.
The model with two independent coupling constants is obtained from $SU(m,n)$ for any $m\geq (n+2)$.

The main results of this letter are the characterization of the reductions of the geodesic system on
a real simple Lie group $G$ under the $G_+\times G_+$
 symmetry presented in Section 2,
where $G$ is non-compact and $G_+$ is  a maximal compact subgroup,
and the derivation of the model (\ref{1.1}) contained in Section 3.
Our derivation of the classical $BC_n$ model should be compared with
the work of Oblomkov \cite{Obl} treating the quantum mechanical
trigonometric $BC_n$ model, in effect,  by quantum Hamiltonian reduction.
See also Section 4 for further discussion.

\section{From free motion to spin Calogero type models}
\setcounter{equation}{0}

Next we briefly recall some group theoretic background material and introduce our notations,
then describe the Hamiltonian reductions
of the free particle on a Lie group to spin Calogero type models in a convenient framework.
The relevant Lie theoretic results are treated in detail in  \cite{Helg,Knapp},
and we refer to \cite{Per,OrtRat} for reviews of
 symplectic geometry and Hamiltonian reduction.

Let $G$ be a non-compact real simple Lie group
with finite centre
and $\G$ its Lie algebra.
Up to conjugation there exists a unique Cartan involution\footnote{
Notationwise  we  pretend that $G$ is  a matrix group.
One may also think of
$\Theta$ concretely as $\Theta(g)= (g^{-1})^\dagger$.}
$\Theta$ of $G$,
for which the associated automorphism $\theta$ of $\G$ induces the
decomposition
\be
\G= \G_+ + \G_-, \qquad \theta(X_\pm) = \pm X_\pm
\quad \forall X_\pm \in \G_\pm,
\label{2.1}\ee
where the restriction of the Killing form $\langle\ ,\ \rangle$ of $\G$
is negative (resp.~positive) definite on $\G_+$ (resp.~on $\G_-$).
The fixed point set of $\Theta$ is a maximal compact subgroup $G_+ \subset G$
with Lie algebra $\G_+$.
The elements of
$\G_-$ are diagonalizable, with real eigenvalues, in the adjoint representation of $\G$
and it is useful to fix a
maximal Abelian subspace $\A\subset \G_-$.
The choice of $\A$ leads to the refined decomposition
\be
\G_- = \A+ \A^\perp, \qquad \G_+ = \M+ \M^\perp,
\label{2.2}\ee
with
\be
\M:= \{ X\in
\G_+\,\vert\, [X,Y]=0\,\,\,\forall Y\in \A\,\}
\label{2.3}\ee
and the complementary spaces $\A^\perp$, $\M^\perp$ defined with the aid of  $\langle\ ,\ \rangle$.
We may write any $X\in\G$ as
$X=X_- + X_+ = X_\A + X_{\A^\perp} + X_\M + X_{\M^\perp}$ according to
(\ref{2.1}) and (\ref{2.2}).
We also need the group corresponding to $\M$, the centralizer of $\A$ in $G_+$,
\be
M:=\{ m\in G_+\,\vert\, m Y m^{-1} =Y \quad \forall Y\in \A\}.
\label{2.4}\ee
We remind in passing that the Weyl group of the Riemannian
symmetric space $G/G_+$ is $W:= \hat M/M$, where
 $\hat M$ is the normalizer of $\A$ in $G_+$.

Let us call an element of $\A$ {\em regular} if its kernel in the adjoint representation
of $\G$ is $\A + \M$.
The set of regular elements, denoted as $\hat \A\subset \A$,  is the union of its
connected components and we choose
an open  Weyl chamber $\check \A \subset \hat \A$ to be such a connected component.
The regular elements of $G$ can be characterized by  admitting a decomposition
of the form
\be
g= g_+ e^q h_+
\qquad
q\in \check \A,\quad
g_+,h_+ \in G_+,
\label{2.5}\ee
and this decomposition is unique up to replacing
$(g_+, h_+)$ by $(g_+ m, m^{-1} h_+ )$ for any $m\in M$.

Denote by $\check G\subset G$ the open dense submanifold  formed
by the regular elements. From now on identify the dual space
$\G^*$ with $\G$ by means of the Killing form
$\langle\ ,\ \rangle$.
Then, using the trivialization defined by right-translations on $G$,
consider the cotangent bundle $T^* \check G$,
\be
P:=T^* \check G \simeq \check G \times \G = \{ (g,J^l)\,\vert\, g\in \check G,\,\,\, J^l\in \G\,\},
\label{2.6}\ee
equipped with the symplectic form $\Omega$ and the Hamiltonian $\H$ of the geodesic system,
\be
\Omega = d \langle J^l, (dg) g^{-1}\rangle,\qquad
\H(g,J^l)= \frac{1}{2} \langle J^l, J^l\rangle.
\label{2.7}\ee
$J^l$ generates the left-translations and $J^r = -g^{-1} J^l g$ generates
the right-translations on $T^*G$.
We shall perform Hamiltonian symmetry reduction relying on
the subgroup $G_+ \times G_+$ of $G\times G$.

To study the reductions of the geodesic system, it is convenient to first extend
it as follows.
Take two arbitrary coadjoint orbits of $G_+$, say  $(\cO^l, \omega^l)$ and $(\cO^r, \omega^r)$.
The orbits are realized as submanifolds of $\G_+ \simeq \G_+^*$ and $\omega^{l,r}$ denote their
own symplectic forms.
The extended system  is  $(P^{\ext}, \H^{\ext}, \Omega^{\ext})$:
\be
P^{\ext}:= P \times \cO^l \times \cO^r=\{ (g,J^l, \xi^l, \xi^r)\,\vert\,
g\in \check G,\, J^l\in \G,\, \xi^l\in \cO^l,\,\xi^r\in \cO^r\,\},
\label{2.8}\ee
\be
\Omega^{\ext} := \Omega + \omega^l + \omega^r,\qquad
\H^{\ext}(g,J^l, \xi^l,\xi^r):= \H(g, J^l).
\label{2.9}\ee
Using the Poisson bracket associated with $\Omega^{\ext}$, the corresponding equation of motion reads
\be
\dot{g} = \{ g, \H^{\ext}\} = J^l g,
\quad
\dot J^l = \{ J^l, \H^{\ext}\} =0,
\quad
\dot \xi^\lambda = \{ \xi^\lambda, \H^{\ext}\} =0\quad \hbox{for}
\quad \lambda=l,r.
\label{2.10}\ee
The solution with initial value $(g(0), J^l, \xi^l,\xi^r)$ yields the geodesic $g(t) =e^{tJ^l} g(0)$.

Now we consider the reduction of the above system based on the symmetry group
$G_+ \times G_+$.
Any $(g_+^l, g_+^r)\in G_+ \times G_+$ operates by the transformation
$T(g_+^l, g_+^r)\in \mathrm{Diff}(P^{\ext})$ defined by
\be
T(g_+^l, g_+^r): \left(g, J^l, \xi^l, \xi^r\right)\mapsto
\left(g_+^l g (g_+^r)^{-1}, g_+^l J^l (g_+^l)^{-1}, g_+^l \xi^l (g_+^l)^{-1}, g_+^r \xi^r (g_+^r)^{-1}\right).
\label{2.11}\ee
The equivariant momentum map, $\Psi = (\Psi^l, \Psi^r): P^{\ext} \to \G_+^* \oplus \G_+^*$,
for this Hamiltonian action is
furnished by
\be
\Psi(g, J^l, \xi^l, \xi^r) = (J^l_+ + \xi^l, -(g^{-1} J^l g)_+ + \xi^r),
\label{2.12}\ee
where the factors $\G_+^*$ are identified with $\G_+$ using the scalar product,
the elements of $\G_+ \oplus \G_+$ are denoted as ordered pairs, and $J^l=J^l_+ + J^l_-$
according to (\ref{2.1}).
We are interested in the reduced Hamiltonian system
$(P_{\red},\Omega_{\red}, \H_{\red})$
obtained  from $(P^{\ext}, \Omega^{\ext}, \H^{\ext})$ at the zero value of the momentum map $\Psi$, i.e.,
\be
P_{\red}:= P^{\ext}_{\Psi=0}/(G_+ \times G_+).
\label{2.14}\ee
It is easy to see that this is equivalent to the (singular)
Marsden-Weinstein  reduction  \cite{OrtRat} of the original system
$(P, \Omega, \H)$ at an arbitrary value $(-\mu^l, -\mu^r)$ of the
corresponding momentum map, $(J^l_+, J^r_+)$, with $\mu^l\in \cO^l$,
$\mu^r\in \cO^r$.
We assume in what follows that $P^{\ext}_{\Psi=0}$
is  {\em non-empty}, which is a
condition on the orbits $\cO^l$, $\cO^r$.
In fact, the condition that $\Psi(g,J^l,\xi^l,\xi^r)=0$ admits a solution on $P^{\ext}$
is equivalent to the consistency of
(\ref{2.20}) below for some
$\xi^l \in \cO^l$ and $\xi^r\in \cO^r$.

Now we are ready to characterize the
reduced Hamiltonian system defined above.
The key step is to utilize that all $G_+\times G_+$ orbits in the constrained manifold
$P^{\ext}_{\Psi=0}$  intersect the following  `gauge slice':
\be
S:= \{ (e^{q}, J^l, \xi^l,\xi^r)\in P^{\ext}_{\Psi=0}\,\vert\, q\in \check \A\,\},
\label{2.15}\ee
since every regular element of $G$ can be transformed into $\exp(\check \A )$
by means of the action (\ref{2.11}).
The gauge slice $S$ represents only a partial
gauge fixing of the gauge transformations defined by the $G_+\times G_+$ action (\ref{2.11}).
The residual gauge transformations
(the maps that transform an arbitrarily chosen point of $S$ into $S$)  are generated by  the subgroup
\be
M_{\diag}:=\{ (m,m) \in G_+\times G_+\,\vert\, m\in M\}.
\label{2.16}\ee
$M_{\diag}$ is naturally isomorphic to, and is below often identified with, $M$.
At this point we arrived at the model
\be
P_{\red}=P^{\ext}_{\Psi=0}/(G_+\times G_+) = S/M_{\diag}.
\label{2.17}\ee
To describe  $P_{\red}$ more explicitly, we use
the  orthogonal complement of the Lie algebra $\M_{\diag} \subset \G_+ \oplus \G_+$ of $M_{\diag}$,
\be
\M_{\diag}^\perp = \{ (X_1,X_2) \in \G_+ \oplus \G_+ \,\vert\, \langle X_1 +X_2, V \rangle =0
\quad \forall V\in \M\,\},
\label{2.18}\ee
with respect to the scalar product $\langle (X_1,X_2), (Y_1,Y_2)\rangle_+ =
 \langle X_1, Y_1\rangle + \langle X_2, Y_2\rangle$ on $\G_+\oplus \G_+$.
By decomposing $J^l\in \G$ and $\xi^\lambda \in \cO^\lambda \subset \G_+$ $(\lambda=l,r)$
according to (\ref{2.2}),
\be
J^l= J^l_\A + J^l_{\A^\perp} + J^l_\M + J^l_{\M^\perp},
\qquad
\xi^\lambda=\xi_\M^\lambda + \xi^\lambda_{\M^\perp},
\label{2.19}\ee
and using that $\ad_q$ ($\forall q\in \check \A$) yields a linear bijection
 between  $\M^\perp$ and $\A^\perp$,
the constraint $\Psi=0$ on $S$ can be solved as follows.
In fact, the condition $\Psi=0$ on $S$  is equivalent to the equations
\be
\xi^l_\M + \xi^r_\M=0,
\label{2.20}\ee
\be
J^l =J^l_\A - F(\ad_q) \xi^l_{\M^\perp} - w(\ad_q)\xi^r_{\M^\perp} - \xi^l,
\label{2.21}\ee
where $J^l_\A\in \A$ is arbitrary and $F$, $w$ are the analytic functions
\be
F(z) = \coth z,
\qquad
w(z) = \frac{1}{\sinh z}.
\label{2.22}\ee
Equation (\ref{2.20}) ensures that $(\xi^l, \xi^r)\in \M_{\diag}^\perp$.
Motivated by the parametrization (\ref{2.21}),
let us introduce the smooth one-to-one map
$I: (\check \A \times \A) \times (\cO^l \oplus \cO^r) \cap \M_{\diag}^\perp \to S$ by
\bea
&&I(q,p, \xi^l, \xi^r):=(e^{q}, \cL(q,p,\xi^l,\xi^r), \xi^l, \xi^r),\nonumber\\
&& \cL(q,p,\xi^l, \xi^r) := p - F(\ad_q) \xi^l_{\M^\perp} - w(\ad_q)\xi^r_{\M^\perp}-\xi^l.
\label{*28}\eea
The pull-back of
$\Omega^{\ext}\vert_S$ by $I$, where $\Omega^{\ext}\vert_S$ is the pull-back of $\Omega^{\ext}$ to
the submanifold $S\subset P^{\ext}$,
turns out to be
\be
I^* (\Omega^{\ext}\vert_S)= d \langle p, dq\rangle + \left(\omega^{l}+
 \omega^r\right)\vert_{(\cO^l \oplus\cO^r) \cap \M_{\diag}^\perp}.
\label{2.25}\ee
The first term is the canonical symplectic structure of
$T^* \check \A \simeq \check \A \times \A = \{ (q,p)\}$.
The second term in (\ref{2.25}) is the restriction of $\omega^l+\omega^r$
to the zero level set of the momentum map for the action of the group
 $M\simeq M_{\diag}$ on $\cO^l\oplus \cO^r$,
provided by $(\xi^l,\xi^r) \mapsto (\xi^l_\M + \xi^r_\M) \in \M\simeq \M^*$.
Notice that $I$ is an $M$ equivariant map,
where $M$ acts trivially on $T^*\check \A$.
On account of its equivariance,
the map $I$  gives rise to the identification
$S/M_{\diag}= T^*\check \A \times \cO_{\red}$ with
\be
\cO_{\red}:= (\cO^l \oplus \cO^r) \cap \M_{\diag}^\perp / M_{\diag}.
\label{2.26}\ee
In terms of the model of $S$  provided by the map $I$ (\ref{*28}), the  Hamiltonian of the geodesic motion
takes the form
\be
(\H^{\ext} \circ I)(q,p, \xi^l,\xi^r) = \frac{1}{2} \langle \cL(q,p,\xi^l,\xi^r),
\cL(q,p,\xi^l,\xi^r)\rangle.
\label{2.27}\ee
By collecting the above formulae and spelling out the
Hamiltonian with the aid of the identity
$F(z) w(z) = \frac{1}{2} w^2(\frac{z}{2}) - w^2(z)$,
we obtain our

\smallskip
\noindent
{\bf Main result:}
{\em
The reduced geodesic system
$(P_{\red}, \Omega_{\red}, \H_{\red})$  defined  above can be identified as
\be
P_{\red} = T^* \check \A \times \cO_{\red},
\qquad
\Omega_{\red}= d \langle p, dq\rangle + \omega_{\red},
\label{2.30}\ee
where $q,p$ are the natural variables on $T^*\check \A$ and
$(\cO_{\red}, \omega_{\red})$ (\ref{2.26}) is the  symplectic reduction of $\cO^l \oplus \cO^r$
by the subgroup $M_{\diag}\subset G_+\times G_+$  at the zero value of its momentum map.
The reduced Hamiltonian yields a hyperbolic spin Calogero type model in general,
since as an $M$ invariant function
on $T^* \check \A \times (\cO^l \oplus \cO^r) \cap \M_{\diag}^\perp$ it has the form
\bea
&& \H_{\red}(q,p, \xi^l,\xi^r)=
 \frac{1}{2}\langle p, p \rangle
-\frac{1}{2}\langle \xi^l_{\M^\perp}, w^2(\ad_q) \xi^l_{\M^\perp} \rangle
-\frac{1}{2}\langle \xi^r_{\M^\perp}, w^2(\ad_q) \xi^r_{\M^\perp} \rangle\nonumber\\
&& \qquad \qquad
+ \frac{1}{2}\langle \xi^l_\M, \xi^l_\M \rangle
+\langle \xi^r_{\M^\perp}, w^2(\ad_q) \xi^l_{\M^\perp} \rangle
-\frac{1}{2} \langle \xi^r_{\M^\perp}, w^2(\frac{1}{2}\ad_q) \xi^l_{\M^\perp} \rangle,
\label{2.31}\eea
where  $w(z) = \frac{1}{ \sinh z}$ and  $\xi^l_\M + \xi^r_\M=0$.
}
\bigskip

Now some remarks are in order.
First, note that our spin Calogero models enjoy
 Weyl group symmetry similarly  to the standard Calogero type models.
This symmetry is not explicit in the above since the   Weyl chambers
are permuted by the Weyl group $W$ and we
have gauge fixed the
coordinate variable $q$ to a single chamber $\check \A$.
However, we could have used in our derivation the larger
 gauge slice, $\hat S$,
 which differs from $S$ (\ref{2.15})
 only in that
 $q$ runs over the full set of  regular elements $\hat \A\subset \A$.
The corresponding residual gauge transformations belong to the normalizer $\hat M$,
and it is easily seen that
$P_{\red}=S/M= \hat S/\hat M = \hat P_{\red}/W$
with
$\hat P_{\red}:= \hat S/M = T^* \hat \A \times \cO_{\red}$.
The point is that the spin Calogero model defined on $\hat P_{\red}$
is invariant
with respect to the natural action of $W=\hat M/M$  induced by the action of
 $\hat M\simeq \hat M_{\diag}\subset G_+\times G_+$ on $\hat S$.

The structure of the reduced phase space described above
 is consistent
with general results on reduced cotangent bundles
derived in \cite{Hoch} under the assumption that only
one isotropy type appears for the action of the
symmetry group on the configuration space.
Indeed,
the isotropy group of any element (\ref{2.5}) of $\check G$  is conjugate
to $M_{\diag}$ for the action of $G_+\times G_+$.
Note that $\cO_{\red}$ (\ref{2.26}) is not a smooth manifold in general.
This does not cause any difficulty, since one can define the smooth functions on $P_{\red}$ (\ref{2.30})
to be the smooth, gauge invariant functions on $P^{\ext}_{\Psi=0}$.
For a review of singular symplectic reduction, see  \cite{OrtRat}.

The solutions of the reduced system $(P_{\red}, \Omega_{\red}, \H_{\red})$ can be obtained
algebraically, by projecting the obvious solution curves of
$(P^{\ext}, \Omega^{\ext}, \H^{\ext})$ (\ref{2.10}) that satisfy the constraint $\Psi=0$.
All spin Calogero models that arise by reduction are integrable in this direct sense.
These models naturally possess many constants of motion, too.
Indeed, $J^\lambda$ and $\xi^\lambda$ ($\lambda=l,r$) are conserved quantities
for the dynamics (\ref{2.10}) on $P^{\ext}$, and any combination of them that is invariant
with respect to the $G_+ \times G_+$ symmetry transformations (\ref{2.11}) induces
a constant of motion for the reduced system.
For example, consider the function on $P_{\red}$ induced by
\be
h (K^\lambda(v))
\quad\hbox{with}\quad
K^\lambda(v):= J^\lambda_- - v \xi^\lambda,
\label{2.33}\ee
where $v$ is any real parameter, $\lambda\in \{l,r\}$, and
$h$ is a $G$ invariant real function on $\G$.
A straightforward calculation, similar to Section 4 in \cite{FP},
 shows that all constants of motion of the form (\ref{2.33})
are in involution on $P_{\red}$.
The Liouville integrability of the reduced systems could be shown
 starting from these remarks.

If desired, one may also construct Lax pairs as follows.
Let $\sigma \subseteq S$ (\ref{2.15}) denote a gauge slice (of a partial or complete
gauge fixing)
and for any $v\in \bR$
define $L^\lambda(v):\sigma \to \G$ by
\be
L^\lambda(v):= K^\lambda(v) \vert_\sigma.
\label{2.34}\ee
With respect to the projection of the Hamiltonian vector field (\ref{2.10}) to $\sigma$,
$L^\lambda(v)$ is found to satisfy a Lax equation
\be
{\dot L}^\lambda(v) = [\y^\lambda, L^\lambda(v)],
\qquad
\lambda=l,r.
\label{2.35}\ee
In fact, proceeding like in \cite{FP} we find that
\bea
&&\y^l= \y_\M + \frac{1}{2} \xi^l_\M - w^2(\ad_q) \xi^l_{\M^\perp} -
(wF)(\ad_q) \xi^r_{\M^\perp},\nonumber\\
&&\y^r= \y_\M + \frac{1}{2} \xi^r_\M - w^2(\ad_q) \xi^r_{\M^\perp} - (wF)(\ad_q) \xi^l_{\M^\perp},
\label{2.36}\eea
where $w$, $F$ appear in (\ref{2.22}) and $\y_\M \in \M$ can be determined
by the consistency of the gauge fixing conditions imposed on $\sigma$.
Equation (\ref{2.10}) also implies that $\dot{q}=p$ and by using this one can verify
that the two Lax equations in (\ref{2.35})
are actually equivalent to each other.

\section{Spinless $BC_n$ Sutherland models from $SU(m,n)$}
\setcounter{equation}{0}

Let us begin by noting that the symmetry reductions based on
$G_+ \times G_+$ can be implemented also as a two step process, say
imposing first the momentum  map constraint
on $J^r_+$. If one chooses $\cO^r=\{0\}$ in this first step, then one obtains the geodesic system on
the symmetric space $G/G_+$, which is  subsequently reduced in the second step imposing
the constraint on $J_+^l$.
The $\cO^r=\{0\}$ special case of the result given by (\ref{2.30}), (\ref{2.31})
 reproduces a result
in \cite{FP}, where we studied the reductions of the geodesic system on $G/G_+$ taking an
arbitrary orbit for $\cO^l$.
In this reference  we also examined the cases for which the reduced phase space is
isomorphic to $T^* \check\A$, which means that the reduced system gives a {\em spinless} Calogero model.
Next we outline a mechanism whereby the models obtained in \cite{FP} can be deformed
whenever $G_+$ admits a one-point coadjoint orbit consisting of a non-zero
(infinitesimal)  character.

Let $C\in \G_+^*\simeq \G_+$ be a non-zero character, i.e.,
 an element invariant under conjugation by $G_+$.
Starting from $\cO_{\red}$ (\ref{2.26}),
we can define a  shifted space of  spin degrees of
freedom   by
\be
\cO_{\red}^y:= \left(\left(\cO^l-yC\right) \oplus \left(\cO^r + yC\right)\right)
\cap \M_{\diag}^\perp / M_{\diag}, \qquad
\forall y\in \bR,
\label{deform}\ee
where $(\cO^r+yC)$ and $(\cO^l-yC)$ are one parameter families of coadjoint orbits of $G_+$.
This is possible since the constraint (\ref{2.20}) is invariant under replacing
$(\xi^l,\xi^r)$ by $(\xi^l - yC, \xi^r + yC)$.
A crucial point to notice is that
if $\cO_{\red}=\cO_{\red}^{y=0}$ is a one-point space, then  this
feature holds for any $y\in \bR$ with
the reduced Hamiltonian  $\H_{\red}$ (\ref{2.31})
{\em acquiring a  dependence  on the `deformation parameter' $y$}.
It is well-known
 \cite{Helg,Knapp} that non-trivial characters exist if and only if $G/G_+$ is a Hermitian
symmetric space, which holds for example if $G=SU(m,n)$, and in these cases the space
of characters is one-dimensional.

In \cite{FP} we explained that one-point reduced orbits (\ref{2.26})  with $\cO^r=\{0\}$ result
if one takes $G=SU(m,n)$ and chooses
$\cO^l$ in a very special manner utilizing  minimal coadjoint orbits of an $SU(k)$ factor of
$G_+$. This is the essential point behind the derivation of
the $BC_n$ Sutherland model from the geodesic system of the
symmetric space of $SU(n+1,n)$ due to Olshanetsky and Perelomov \cite{OPInv,OPCim,OPRep1}.
However, the three coupling constants of the model
resulting from their procedure are necessarily subject
to a quadratic relation.
Here, we utilize the one parameter family of characters of $G_+$
to increase the number of independent coupling constants in the reduced Hamiltonian by one.
In fact, we show below that in this way the classical $BC_n$
Sutherland model with three independent coupling
constants can be obtained as a reduction of the geodesics on $SU(n,n)$ and on $SU(n+1,n)$.

We need some further notations.
Consider the joint eigensubspaces of the elements of $\A$,
\be
\G_\alpha:= \{
X\in \G\,\vert\, [Y,X]=\alpha(Y)X\,\,\,\forall Y\in \A\,\}.
\label{3.1}\ee
The linear
functions $\alpha\in \A^*\setminus\{0\}$  with $\dim (\G_\alpha) \neq 0$  are called
restricted roots.
They form a crystallographic root system, denoted by $\cR$.
The subspaces in (\ref{2.2}) satisfy
$\M^\perp + \A^\perp = \oplus_{\alpha\in \cR} \G_\alpha$.
We fix a polarization $\cR = \cR_+ \cup
\cR_-$ and choose  $E_\alpha^a\in \G_\alpha$
($a=1,\ldots, \nu_\alpha:= \mathrm{dim}(\G_\alpha)$) so that
\be
\theta(E_\alpha^a) = - E_{-\alpha}^a, \qquad \langle E_\alpha^a,
E_\beta^b\rangle = \delta_{\alpha, -\beta} \delta^{a,b}.
\label{3.2}\ee
Then $\M^\perp$ and $\A^\perp$ are spanned by
\be
E_\alpha^{+,a} = \frac{1}{\sqrt{2}} ( E_\alpha^a +
\theta(E_\alpha^a)) \in \M^\perp, \qquad E_\alpha^{-,a} =
\frac{1}{\sqrt{2}} ( E_\alpha^a - \theta(E_\alpha^a)) \in \A^\perp
\qquad \forall \alpha\in \cR_+.
\label{3.3}\ee
Let us now focus on  $SU(m,n)$ and its Lie algebra $su(m,n)$,  given by
\bea
&& SU(m,n)=\{g\in SL(m+n,\mathbb{C})\,\vert\, g^\dagger I_{m,n}g=I_{m,n}\},
\label{3.4}\\
&& su(m,n)=\{X\in\ sl(m+n,\mathbb{C})\,\vert\, X^\dagger I_{m,n}
+I_{m,n}X=0\},
\label{3.5}
\eea
where $I_{m,n}:=\mathrm{\diag}(\1_m,  -\1_n)$, $m\geq n$ and $\1_k$ ($k=m,n$) is the
$k\times k$ identity matrix.
A block matrix  $X\in \G=su(m,n)$ reads
\be
X=\left(\begin{array}{cc}
A & B\\
B^\dagger & D
\end{array}\right),
\label{3.6}\ee
where $B\in\mathbb{C}^{m\times n}$, $A\in u(m)$, $D\in u(n)$ and $\tr A + \tr D=0$.
The Cartan involution of $G=SU(m,n)$ is
$\Theta: g \mapsto (g^\dagger)^{-1}$.
Thus
\be
G_+ = S( U(m) \times U(n)),
\label{3.7}\ee
\be
\G_+= su(m) \oplus su(n) \oplus \bR C_{m,n}=
\left\{\left(\begin{array}{cc}
A & 0\\
0 & D
\end{array}\right) + x C_{m,n}\Bigg|\:
A\in su(m),\, D\in su(n),\, x\in \bR \right\}
\label{3.8}\ee
with the central element
\be
C_{m,n}:=\mathrm{\diag}( \ri n \1_m, -\ri m\1_n),
\label{3.9}\ee
which spans the space of characters.
A  maximal Abelian subspace of $\G_-$ is furnished by
\be
\A:=\left\{
\,q:=\left(\begin{array}{ccc}
\0_n & 0 & Q\\
0 & \0_{m-n} & 0\\
Q & 0 & \0_n
\end{array}\right)
 \:\Bigg|\:
Q=\diag(q^1,\ldots,q^n),\: q^j\in\mathbb{R}\right\}.
\label{3.10}\ee
Taking $\chi:=\diag(\chi_1,\ldots,\chi_n)$ with any $\chi_j\in\mathbb{R}$,
the centralizer of $\A$ in $\G_+$ is
\be
\M=\{\diag(\ri\chi,\gamma,\ri\chi) \: |\:
\gamma\in u(m-n),\,
\tr \gamma + 2\ri \tr\chi =0\},
\label{3.11}\ee
and the subgroup $M$ of $G_+$ is
\be
M=\{ \diag( e^{\ri \chi}, \Gamma, e^{\ri\chi}) \: | \:   \Gamma \in U(m-n),\,
(\det \Gamma)(\det e^{\ri 2\chi}) =1 \}.
\label{3.12}\ee
One may define the functionals $e_k\in \A^*$ ($k=1,\ldots, n$) by
$e_k(q):= q^k$.
The system of restricted roots is of $BC_n$ type if $m>n$ and of
$C_n$ type if $m=n$.
Indeed, we have
\be
\cR_+:=\{e_j\pm e_k\:(1\leq j<k\leq n),\, 2e_k,  e_k\:(1\leq k\leq n)\}
\quad \hbox{if}\quad m>n,
\label{3.13}\ee
\be
\cR_+:=\{e_j\pm e_k\:(1\leq j<k\leq n),\, 2e_k \:(1\leq k\leq n)\}
\quad \hbox{if}\quad m=n,
\label{3.14}\ee
with the multiplicities
\be
\nu_{e_j \pm e_k}= 2 \quad (1\leq j<k\leq n),\quad
\nu_{2e_k}= 1\quad\hbox{and}\quad
\nu_{e_k}= 2(m-n) \quad (1\leq k\leq n).
\label{3.15}\ee
We  adopt the  convention  described explicitly in \cite{FP},
where the basis vectors of $\M^\perp$ (\ref{3.3}) are denoted as
\be
E^{+,\rr}_{e_j\pm e_k},
\quad
E^{+,\ri}_{e_j\pm e_k},
\quad
E^{+,\ri}_{2e_k},
\quad
E^{+,\rr,d}_{e_k},
\quad
E^{+,\ri,d}_{e_k}
\quad\hbox{for}\quad
1\leq d\leq (m-n).
\label{3.16}\ee
The superscripts $\rr$ or $\ri$ refer to purely real or imaginary matrices.

Since $G_+$ (\ref{3.7}) contains factors of $SU(k)$ type, we can use
the minimal coadjoint orbits of $SU(k)$
in our reduction procedure, which  underlie also the derivation \cite{KKS}  of
the $k$-particle Sutherland model from the geodesic motion on $SU(k)$.
For any $u\in \bC^k$, viewed as a column vector, we define
\be \eta_\pm (u):= \pm \ri
\left( u u^\dagger - \frac{u^\dagger u}{k} {\bf 1}_k\right)\in su(k).
\label{3.17}\ee
The minimal coadjoint orbits of $SU(k)$ are provided by
\be
\cO_{k,\kappa, \pm}:= \{\xi \in su(k)  \, \,\vert\, \exists u \in \bC^k,\,\, u^\dagger
u = k \kappa,\,\, \xi = \eta_\pm(u) \,\},
\label{3.18}\ee
where $\kappa>0$ is a constant.
For definiteness,
we below take the plus sign.

For $G=SU(n,n)$, we now consider the following coadjoint orbits of $G_+$:
\be
\cO^l:= \cO_{n,\kappa,+} + \{ x C_{n,n}\},
\qquad
\cO^r := \{ y C_{n,n}\},
\label{3.19}\ee
where $x$ and $y$ are real constants
and $\cO_{n,\kappa, +}$ is embedded say in the upper $su(n)$ block of
$\G_+$.
Since $C_{n,n}\in \M^\perp$,  no restriction on $x$, $y$, $\kappa$ arises from the constraint (\ref{2.20}).
One may confirm in the standard manner \cite{KKS,FP} that the reduced orbit  $\cO_{\red}$
(\ref{2.26})
consists of a single point, and as a representative  one can take
\be
\xi^l:= \kappa \sum_{1\leq j < k\leq n}
\left(E^{+,\ri}_{e_j + e_k} + E^{+,\ri}_{e_j - e_k}\right) + \sqrt{2} x n \sum_{k=1}^n E^{+,\ri}_{2e_k},
\quad
\xi^r := yC_{n,n} =\sqrt{2} y n \sum_{k=1}^n E^{+,\ri}_{2e_k}.
\label{3.20}\ee
Upon substitution into (\ref{2.31}) using the normalization (\ref{3.2}), $\langle X,Y\rangle := \tr(XY)$,
the reduced Hamiltonian (\ref{2.31}) now gives
\be
\frac{1}{2} \H^{SU(n,n)}_{\red}(q,p,\xi^l,\xi^r) =\H_{BC_n}(q,p)
\quad\hbox{with}\quad
g^2 = \frac{\kappa^2}{4},
\quad
g_1^2 = \frac{x y n^2}{2},
\quad
g_2^2 = \frac{(x-y)^2 n^2}{2},
\label{3.22}\ee
where we use the  notation (\ref{1.1}).
The
coupling constants $g^2$, $g_1^2$, $g_2^2$
can take arbitrary positive values, and we may even change the sign of $g_1^2$
by changing the sign of $xy$.
This association of the classical $BC_n$ Sutherland model with $SU(n,n)$ appears to be  a new result.
By setting $y=0$ we reproduce the $C_n$ type Hamiltonian previously known
to arise from $SU(n,n)$ \cite{OPRep1,ABT} and $x=y\neq 0$
(resp.~$x=y=0$) yields the $B_n$ (resp.~$D_n$) type Sutherland Hamiltonian.

For $G=SU(n+1,n)$, we take  $\cO^l$ and $\cO^r$ to be
\be
\cO^l:= \cO_{n+1,\kappa,+} + \{ x C_{n+1,n}\},
\qquad
\cO^r := \{ y C_{n+1,n}\},
\label{3.23}\ee
where $\cO_{n+1,\kappa, +}$ is embedded into the $su(n+1)$ factor of $\G_+$.
An analysis similar to \cite{FP} shows that the consistency of the constraint (\ref{2.20}) requires
\be
\kappa + x + y \geq 0
\quad\hbox{and}\quad
\kappa - n(x+y) \geq 0.
\label{3.24}\ee
The reduced orbit (\ref{2.26}) again consists of a single point, and for a representative one can use
\be
\xi^l := - \xi^r_\M +
2 g \sum_{1\leq j < k\leq n}
\left(E^{+,\ri}_{e_j + e_k} + E^{+,\ri}_{e_j - e_k}\right) +
2 h_1  \sum_{k=1}^n E^{+,\ri,1}_{e_k}
+2 h_2  \sum_{k=1}^n E^{+,\ri}_{2e_k},
\label{3.25}\ee
\be
g=\frac{\kappa + x +y}{2},\quad
h_1= \frac{\sqrt{(\kappa + x+ y) ( \kappa -nx -ny)}}{\sqrt{2}},
\quad
h_2 =\frac{2(n+1)x +y}{\sqrt{8}},
\label{3.26}\ee
\be
\xi^r_\M =-\frac{\ri y}{2} \diag(\1_n, - 2n, \1_n),
\quad
\xi^r_{\M^\perp}= 2 \tilde h_2  \sum_{k=1}^n E^{+,\ri}_{2e_k},
\quad
\tilde h_2 = \frac{ y ( 2n+ 1)}{\sqrt{8}}.
\label{3.27}\ee
Referring to $\H_{BC_n}$ in (\ref{1.1}), in the present case we find
\be
\frac{1}{2} \H^{SU(n+1,n)}_{\red}(q,p,\xi^l,\xi^r) =\H_{BC_n}(q,p)
-\frac{ y^2 (2 n^2 + n)}{8}
\quad\hbox{with}\quad
g_1^2 = h_1^2 + h_2 \tilde h_2,
\quad
g_2^2 = (h_2 - \tilde h_2)^2.
\label{3.28}\ee
The coupling constants $g^2$, $g_1^2$, $g_2^2$ of $\H_{BC_n}$ depend
on the three parameters $x$, $y$, $\kappa$ subject to (\ref{3.24}),
and one recovers the result of \cite{OPInv,FP} upon setting $y=0$.

In the above we have seen how the spinless $BC_n$ Sutherland model with three
arbitrary coupling constants arises from
$SU(n,n)$ and from $SU(n+1,n)$.
What happens if $m\geq (n+2)$?
Briefly,
in these cases we can obtain a one-point reduced orbit $\cO_{\red}$ (\ref{2.26}) if
\be
\cO^l= \cO_{n,\kappa,+} + \{ x C_{m,n}\},
\qquad
\cO^r = \{ y C_{m,n}\},
\qquad
 x=-y.
\label{3.30}\ee
The orbit $\cO_{n,\kappa,+}$ is embedded in the $su(n)$ factor of $\G_+$ (\ref{3.8}).
The condition $(x+y)=0$ is now enforced by the constraint (\ref{2.20}).
This leads again to the $BC_n$ model, but with only two independent coupling parameters.
Concretely, we find that
\be
 \frac{1}{2}\H^{SU(m,n)}_{\red} = \H_{BC_n}
-\frac{y^2 (m^2-n^2) n}{8}
\quad\hbox{with}\quad
g^2 = \frac{\kappa^2}{4},\quad
g_1^2 = -\frac{g_2^2}{4}= -\frac{y^2 (m+n)^2}{8}.
\label{3.21}\ee
In the $y=0$ case \cite{FP}  the model $\H^{SU(m,n)}_{\red}$  becomes of type $D_n$.
Finally, we note that the choice (\ref{3.30}) is available for $m=n+1$ as well.

One can spell out the Lax matrices (\ref{2.34})
for all the above cases and can also determine the explicit form of
$\y_\M$ in (\ref{2.36}).
The Lax pairs derived in this way appear to be closely related to the
Lax pairs of the $BC_n$ model (\ref{1.1})
obtained in \cite{InoMes} by a different method.

\section{Discussion}
\setcounter{equation}{0}

The main results of this letter are the general description of the reduced
geodesic system $(P_{\red},\Omega_{\red}, \H_{\red})$ presented in Section 2
and the realization that this contains  the spinless $BC_n$ Sutherland
models (\ref{1.1}) with three independent coupling constants as explained in Section 3.
The results can be extended to compact Lie
groups straightforwardly, in correspondence with
the trigonometric version of the hyperbolic (spin) Calogero models encoded by (\ref{2.31}).

It could be interesting to investigate  generalizations
based on replacing the group $G_+ \times G_+$
by suitable groups $G_+' \times G_+''$, where the factors
are fixed by two commuting involutions of $G$.
One can proceed as in  Section 2 whenever
a  `good decomposition' analogous to (\ref{2.5}) is available.

The models (\ref{2.31})
can be quantized by quantum Hamiltonian reduction as follows.
One starts by replacing the coadjoint orbits $\cO^\lambda$ in (\ref{2.8})
by
irreducible unitary representations $\rho_\lambda$ of $G_+$ on vector
spaces $V_\lambda$ for $\lambda=l,r$ and considers  also the associated
representation $\rho$ of $G_+ \times G_+$ on $V= V_l \otimes V_r$.
The quantum analogue of  $P^{\ext}$ (\ref{2.8}) is
the Hilbert space of $V$ valued square integrable wave functions
on $\check G$ and quantum Hamiltonian reduction amounts to
allowing only those wave functions $\psi$ that are
equivariant in the sense that $\psi(g_+^l g (g_+^r)^{-1})=
\rho(g_+^l, g_+^r) \psi(g)$ holds.
Because of the equivariance propery,
these functions are determined by their restrictions to the
domain $\exp(\check \A)$ and
the restricted wave functions take their values
in the subspace $V^M$ of $V$  spanned by the vectors invariant
under the subgroup $M_{\diag}$ of $G_+\times G_+$.
The allowed representations must therefore satisfy
$\dim(V^M) >0$. Spinless Calogero type models arise at the quantum mechanical
level if $\dim(V^M)=1$.
The reduced Hilbert space
naturally comes
equipped with a commuting family of self-adjoint operators induced
by the centre of the universal enveloping algebra of $\G$.
This perspective on quantum Calogero type models
originates from \cite{OPfa}, where
the trivial representation was taken for the $\rho_\lambda$ above.
Many interesting results obtained in this framework can be found in
 \cite{OPRep2, Heck, Eting, Poly, EFK} and references therein.

We plan to elaborate the consequences of the quantum Hamiltonian
reduction in a future publication, where we shall also deal with
the relationship between
the reduction procedure proposed in Section 3 and the interpretation
of $BC_n$ type  Jacobi  polynomials
as generalized spherical functions
on $GL(m+n, \bC)/ (GL(m,\bC) \times GL(n,\bC))$ put forward by Oblomkov \cite{Obl}.
It is well-known  (see e.g.~\cite{Heck}) that these polynomials
give the eigenstates of the $BC_n$
type trigonometric Sutherland model.
However,
the natural compact analogue of Oblomkov's construction,
obtained by substituting  $SU(m+n)$ for $GL(m+n,\bC)$,
does not seem to
coincide with the quantized version of our classical Hamiltonian reduction,
except in the $m=n$ case.
For $m\neq n$, his construction and ours may produce
the eigenstates of the $BC_n$ Hamiltonian for different discrete sets of the coupling constants, but
it requires further work to clarify this issue.

\bigskip
\noindent{\bf Acknowledgements.}
The work of L.F. was supported in part by the Hungarian
Scientific Research Fund (OTKA) under grants
T043159, T049495  and by the EU networks `EUCLID'
(HPRN-CT-2002-00325) and `ENIGMA'
(MRTN-CT-2004-5652).
He wishes to thank M.A. Olshanetsky  for discussions.
B.G.P. is grateful to J. Harnad for
hospitality in Montreal.


\begin{thebibliography}{99}

\bibitem{Cal}
F. Calogero,
{\it Solution of the one-dimensional $N$-body problem with quadratic and/or
inversely quadratic pair potentials,}
J. Math. Phys. 12 (1971) 419-436.

\bibitem{OPRep1}
M.A. Olshanetsky and A.M. Perelomov,
{\it Classical integrable finite-dimensional systems related to Lie algebras},
Phys. Rept. 71 (1981) 313-400.

\bibitem{OPRep2}
M.A. Olshanetsky and A.M. Perelomov,
{\it Quantum integrable systems related to Lie algebras},
Phys. Rept. 94 (1983) 313-404.

\bibitem{Per}
A.M. Perelomov,
Integrable Systems of Classical Mechanics and Lie Algebras,
Birkh\"auser, 1990.

\bibitem{Heck}
G. Heckman,
{\it Hypergeometric and spherical functions},
pp. 1-89 in:
G. Heckman and H. Schlichtkrull,
Harmonic Analysis and Special Functions on Symmetric Spaces,
Perspectives in Mathematics 16, Academic Press, 1994.


\bibitem{Nekr}
N. Nekrasov,
{\it Infinite-dimensional algebras, many-body systems and gauge theories},
pp. 263-299 in: Moscow Seminar in Mathematical Physics, AMS Transl. Ser. 2,
A.Yu. Morozov and M.A. Olshanetsky (Editors), Amer. Math. Soc., 1999.

\bibitem{dHP}
E. D'Hoker and D.H. Phong,
{\it Seiberg-Witten theory and Calogero-Moser systems},
Prog. Theor. Phys. Suppl. 135 (1999) 75-93, hep-th/9906027.

\bibitem{Diej}
J.F. van Diejen and L. Vinet (Editors), Calogero-Moser-Sutherland Models,
Springer, 2000.

\bibitem{SutWSci}
B. Sutherland,
Beautiful Models, World Scientific, 2004.

\bibitem{Finkel}
F. Finkel, D. G\'omez-Ullate, A. Gonz\'alez-L\'opez, M.A. Rodr\'\i guez and R. Zhdanov,
{\it A survey of quasi-exactly solvable systems and spin Calogero-Sutherland models},
pp. 173-186 in:  Superintegrability in Classical and Quantum Systems,
P. Tempesta et. al. (Editors), Amer. Math. Soc., 2004.

\bibitem{Eting}
P. Etingof,
{\it Lectures on Calogero-Moser systems},
math.QA/0606233.

\bibitem{Poly}
A.P. Polychronakos,
{\it Physics and mathematics of Calogero particles},
J. Phys. A: Math. Gen. 39 (2006) 12793-12845,
hep-th/0607033.

\bibitem{AniJev}
I. Aniceto and A. Jevicki,
{\it Notes on collective field theory of matrix and spin Calogero models},
J. Phys. A: Math. Gen. 39 (2006) 12765-12791,
hep-th/0607152.

\bibitem{OPInv}
M.A. Olshanetsky and A.M. Perelomov,
{\it Completely integrable Hamiltonian systems connected with semisimple Lie algebras,}
Invent. Math. 37 (1976) 93-108.

\bibitem{OPCim}
M.A. Olshanetsky and A.M. Perelomov,
{\it Explicit solutions of some completely integrable systems,}
Lett. Nuovo Cim. 17 (1976) 97-101.

\bibitem{InoMes}
V.I. Inozemtsev and D.V. Meshcheryakov,
{\it Extension of the class of integrable dynamical systems
connected with semisimple Lie algebras},
Lett. Math. Phys. 9 (1985) 13-18.

\bibitem{Opd}
E.M. Opdam,
{\it Root systems and hypergeometric functions IV},
Compositio Math. 67 (1988) 191-209.

\bibitem{OshSek}
T. Oshima and H. Sekiguchi,
{\it Commuting families of differential operators invariant under
the action of a Weyl group},
J. Math. Sci. Univ. Tokyo 2 (1995) 1-75.


\bibitem{Sas}
A.J. Bordner, R. Sasaki and K. Takasaki,
{\it Calogero-Moser models II: symmetries and foldings},
Prog. Theor. Phys. 101 (1999) 487-518,
hep-th/9809068.

\bibitem{FP}
L. Feh\'er and B.G. Pusztai,
{\it Spin Calogero models associated with Riemannian symmetric
spaces of negative curvature},
Nucl. Phys. B 751 (2006) 436-458,
math-ph/0604073.

\bibitem{Obl}
A. Oblomkov,
{\it Heckman-Opdam's Jacobi polynomials for the
 $BC_n$ root system and generalized spherical functions},
 Adv. Math. 186 (2004) 153-180, math.RT/0202076.

\bibitem{Helg}
S. Helgason,
Differential Geometry, Lie Groups, and Symmetric Spaces,
Academic Press, 1978.

\bibitem{Knapp}
A.W. Knapp,
Lie Groups Beyond an Introduction, Progress in Mathematics 140, Birkh\"auser, 2002.

\bibitem{OrtRat}
J.-P. Ortega and T.S. Ratiu,
Momentum Maps and Hamiltonian Reduction,
Progress in Mathematics 222, Birkh\"auser, 2004.

\bibitem{Hoch}
S. Hochgerner,
{\it Singular cotangent bundle reduction and spin Calogero-Moser systems,}
math.SG/0411068.

\bibitem{KKS}
D. Kazhdan, B. Kostant and S. Sternberg,
{\it Hamiltonian group actions and dynamical systems of Calogero type,}
Comm. Pure Appl. Math. XXXI (1978) 481-507.

\bibitem{ABT}
J. Avan, O. Babelon and M. Talon,
{\it Construction of the classical $R$-matrices for the Toda and Calogero models,}
Alg. and Anal. 6 (1994) 67-89, hep-th/9306102.


\bibitem{OPfa}
M.A. Olshanetsky and A.M. Perelomov,
{\it Quantum systems related to root systems, and radial parts of Laplace operators},
Funct. Anal. Appl. 12  (1978) 121-128, math-ph/0203031.


\bibitem{EFK}
P.I. Etingof, I.B. Frenkel and A.A. Kirillov Jr.,
{\it Spherical functions on affine  Lie groups},
Duke Math. J. 80 (1995) 59-90, hep-th/9407047.


\end{thebibliography}
\end{document}